\documentclass[12pt]{iopart}
\usepackage{iopams}
\usepackage{overpic}

\begin{document}

\title[Max-min-plus expressions for 1D particle CA obtained from a fundamental
diagram]{Max-min-plus expressions for one-dimensional particle cellular
automata obtained from a fundamental diagram}

\author{Takazumi Okumura$^1$, Junta Matsukidaira$^1$ and Daisuke Takahashi$^2$}

\address{$^1$ Department of Applied Mathematics and Informatics, Ryukoku
University 1-5, Yokotani, Seta Oe-cho, Otsu, Shiga 520-2194, Japan}
\address{$^2$ Major in Pure and Applied Mathematics, Waseda University,
3-4-1 Okubo, Shinjuku-ku, Tokyo 169-8555, Japan}
\ead{junta@rins.ryukoku.ac.jp}

\begin{abstract}
We study one-dimensional neighborhood-five conservative cellular
automata (CA), referred to as particle cellular automata five (particle
CA5). We show that evolution equations for particle CA5s that belong to certain
types can be obtained in the form of max-min-plus expressions from a
fundamental diagram. The obtained equations are transformed into other
max-min-plus expressions by ultradiscrete Cole-Hopf transformation,
which enable us to analyze the asymptotic behaviors of general
solutions. The equations in the Lagrange representation, which describe
particle motion, are also presented, which too can be obtained from a
fundamental diagram. Finally, we discuss the generalization to a
one-dimensional conservative neighborhood-$n$ CA, i.e., particle CA$n$.
\end{abstract}
\submitto{\JPA}
\maketitle
\section{Introduction}
Cellular automata (CA) are dynamic systems in which space and time are
discrete and physical quantities take on a finite set of discrete
values. Despite their simple construction, CA exhibit
complicated behavior and generate complex patterns. By virtue of
these properties, CA have been used as mathematical models for complex
phenomena in the fields of physics, chemistry, biology, economics, and
sociology\cite{Wolfram,Wolfram2}.

Recently, one-dimensional conservative CA have attracted much
attention as models for traffic flow\cite{Boccara-Fuks}. One of the
simplest models is Rule 184, which is a neighborhood-three CA (also
called an elementary CA). Since Rule 184 exhibits phase transition from
a free-flow state to a congestion state as real traffic flow does, it has
been used as a basic model and extended to many other CA
models. However, despite the intensive research on CA models for traffic
flow, there are few studies dealing with
all one-dimensional conservative CA in a unified way.

A powerful method for dealing with CA is the
ultradiscretization method\cite{TTMS,MSTTT}, which connects difference equations and CA
by applying the following simple formula for the transformation of
variables:
\begin{equation}
 \lim_{\varepsilon\rightarrow
  +0}\varepsilon\log\left(e^{A/\varepsilon}+e^{B/\varepsilon}\right) =
  \max (A,B).\label{udlimit}
\end{equation}
A conservative CA called the "box and ball system" (BBS)
is obtained from the discrete KdV equation using this method, and other
CA that possess a solitonic nature are obtained from soliton
equations. Moreover, Rule 184 is derived from the discrete Burgers
equation\cite{Nishinari-Takahashi}.

Equation (\ref{udlimit}) shows that the addition of the difference equations
corresponds to the max operation of the ultradiscrete equation. The
multiplication and division correspond to $+$ and $-$, respectively, since
we have
\[
 \varepsilon\log\left(e^{A/\varepsilon}\times e^{B/\varepsilon}\right) =
  A + B,\qquad  \varepsilon\log\left(e^{A/\varepsilon}/ e^{B/\varepsilon}\right) = A - B
\]
Though the subtraction is not well defined for the ultradiscrete
equation, we can automatically obtain the ultradiscrete equation and its
solution by replacing $+$, $\times$ and $/$ by
max, $+$ and $-$, respectively, if the subtraction is not included
explicitly in the difference equations and their solutions. The algebra
defined by operations of max, $+$, and $-$ is called "max-plus"
algebra. Binary operations such as AND, OR, and NOT can be easily
converted to max-plus operations.

In a previous study\cite{Takahashi-Matsukidaira-Hara-Feng}, we
investigated one-dimensional neighborhood-four conservative CA, referred
to as particle cellular automata four (particle CA4), and found that
evolution equations for particle CA4-1, 4-2, and 4-3 can be obtained in
the form of max-plus expressions in combination with ``min,'' i.e.,
``max-min-plus'' expressions, which are the key to solving the equations
and analyzing asymptotic behaviors.

In this study, we investigate particle CA5s and show that the evolution
equations for particle CA5s that belong to certain types can be obtained
in the form of max-min-plus expressions from fundamental diagrams. In
addition, we show that the equations obtained are transformed into other
max-min-plus expressions by ultradiscrete Cole-Hopf transformation,
which enables us to analyze the asymptotic behaviors of general
solutions. Furthermore, we give the Lagrange representation for the
class of particle CA5s, which describes the motion of particles and can
be obtained from a fundamental diagram. Finally, we discuss the
generalization to the neighborhood-$n$ CA case, that is, particle CA$n$.

\section{Particle CA}
Let us consider the following equation
\begin{equation}
 u_j^{n+1} = f(u_{j-l}^n, u_{j-l+1}^n, \ldots, u_{j+r}^n) \qquad
  (-\infty < j < \infty, 0 \le n), \label{R-neighbor CA}
\end{equation}
where $n$ and $j$ are integers representing the timestep and space
site number, respectively, and $l$ and $r$ are positive integer
constants. Assuming that $u$ takes the value of 0 or 1, (\ref{R-neighbor
CA}) is an evolution equation for a one-dimensional CA
with the rule defined by $R (=l+r+1)$ variable function $f$. We
call this CA a "neighborhood-$R$ CA." Following
Wolfram\cite{Wolfram,Wolfram2}, to each Rule $f$, we assign a rule
number $N(f)$ such that
\[
 N(f) = \sum_{(u_1, u_2, \ldots,u_R)\in\{0,1\}^R}f(u_1, u_2, \ldots, u_R)2^{2^{R-1}u_1+2^{R-2}u_2+\ldots+2^0u_R}.
\]

In this study, we focus on the one-dimensional CAs that satisfy
\begin{equation}
 \sum_{j=1}^K u_j = \sum_{j=1}^K f(u_{j-l}, u_{j-l+1}, \ldots,
  u_{j+r}) \label{PCA}
\end{equation}
where $K (\ge R)$ is the size of the one-dimensional lattices, and
$u_{j+K}=u_j$ holds for all $j$. From (\ref{PCA}), the condition
\begin{equation}
 \sum_{j=1}^K u_j^n = \sum_{j=1}^K u_j^{n+1}, \label{PCAcond}
\end{equation}
holds, which means that the sum of $u$, that is the number of 1s at all
space sites, is conserved for arbitrary $n$. Let $u_j^n$ denote the
number of particles at the $j$th site and the $n$th timestep, and each "1" in
the solution represents a particle. Then, particles move among sites
according to the evolution rule defined by (\ref{R-neighbor CA}) without creation or
annihilation. We call the CA satisfying this condition (\ref{PCAcond}) a
"particle CA" in this sense.
\section{Particle CA5}
In \cite{Hattori-Takesue}, Hattori and Takesue showed that a
neighborhood-$R$ CA with Rule $f$ is a particle CA if and only if $f$ satisfies
\begin{eqnarray}
 f(u_1,\, u_2,\, \ldots,\, u_R) - u_{l+1} = q(u_1,\, u_2,\, \ldots,\,
  u_{R-1}) - q(u_2,\, u_3,\, \ldots,\, u_R),\label{HT1} \\
 q(u_1,\, u_2,\, \ldots,\, u_{R-1}) =
 \sum_{k=1}^l u_k -\sum_{k=1}^{R-1}f(\underbrace{0,\, 0,\,\ldots,0}_k,\, u_1,\, u_2,\,
 \ldots,\, u_{R-k}),\label{HT2}
\end{eqnarray}
for all $(u_1,u_2,\ldots,u_R)\in\{0,1\}^R$. 

Applying Hattori and Takesue's result to neighborhood-five CA expressed
by the equation
\[
 u_j^{n+1} = f(u_{j-2}^n, u_{j-1}^n, u_j^n, u_{j+1}^n, u_{j+2}^n), 
\]
we obtain a system of $2^5$ linear equations with $2^5$ variables
$f(a, b, c, d, e)$ for all $(a, b, c, d, e)\in\{0,1\}^5$
from (\ref{HT1}) and (\ref{HT2}). By finding binary solutions to the
system, we obtain 428 rules for one-dimensional conservative
neighborhood-five CA (particle CA5). Reflection symmetry
\[
 f_1(a, b, c, d, e) = f_2(e, d, c, b, a) 
\]
and Boolean conjugation symmetry
\[
 f_1(a, b, c, d, e) = 1 - f_2(1-a, 1-b, 1-c, 1-d, 1-e) 
\]
for rules $f_1$ and $f_2$ reduce the 428 rules to 129
rules. Furthermore, by excluding the rules for neighborhood-three and
-four CA (particle CA3 and particle CA4, respectively), we have the following 115 rules:
\bigskip
\begin{center}
\parbox{14cm}{
2863377064, 2881005752, 2881267852, 2881398914, 2881464448, 2944969912,
2945232012, 2945363074, 2945428608, 2947326124, 2947457186, 3098065832,
3099375756, 3099506818, 3099572352, 3102247912, 3103295736, 3103557836,
3103688898, 3103754432, 3116153216, 3120335296, 3132144268, 3132275330,
3136326348, 3136457410, 3136522944, 3137047048, 3148921728, 3153103808,
3153627912, 3163077816, 3163470978, 3163536512, 3165434028, 3165630624,
3167259896, 3167521996, 3167653058, 3167718592, 3169616108, 3169812704,
3180117376, 3182211488, 3184299456, 3186393568, 3196108428, 3198202540,
3200290508, 3200487104, 3201011208, 3202384620, 3202581216, 3203105320,
3213933712, 3214980000, 3216027824, 3216289924, 3217067968, 3217592072,
3218115792, 3218639896, 3219162080, 3219686184, 3220209904, 3220472004,
3220734008, 3220996108, 3221127170, 3366517672, 3367565496, 3367827596,
3370699752, 3372009676, 3384605056, 3388787136, 3400596108, 3404778188,
3421555648, 3422079752, 3431529656, 3431791756, 3435711736, 3448569216,
3450663328, 3452751296, 3454845408, 3464560268, 3482385552, 3484479664,
3484741764, 3486043912, 3486567632, 3487091736, 3487613920, 3488138024,
3488923844, 3489185848, 3639663552, 3640187656, 3703627712, 3704151816,
3705199640, 3705721824, 3706245928, 3706769648, 3707031748, 3707293752,
3771264248, 3822120144, 3822644248, 3824214256, 3824738360, 3888178416,
4040228048,
}
\end{center}
\medskip
\noindent
which is the smallest rule number among their equivalent
rules. In this paper, we assign a number "$m$" in the range 1-115 to each rule,
and call them particle CA5-$m$.

\section{Obtaining evolution equations in the form of max-min-plus expressions
from a fundamental diagram} In the field of traffic-flow theory,
a fundamental diagram is a useful tool for determining the traffic state of a
roadway. This diagram relates traffic flux
and traffic density. Under periodic boundary conditions, the density
is defined by
\[
 \rho = \frac{1}{K}\sum_{j=1}^K u_j^n, 
\]
where $K$ is the period of sites. Since the number of particles is
conserved, $\rho$ is a constant irrespective of time $n$. The average flux
is defined by
\[
 \bar{q}^n = \frac{1}{K}\sum_{j=1}^K q(u_j^n,\ldots,u_{j+R-1}^n). 
\]
For large enough $n$, the evolution often approaches a steady state where $\bar{q}^n$ becomes a constant. In case of convergence, the
constant is defined by
\[
 Q = \lim_{n\rightarrow\infty}\bar{q}^n. 
\]
The graph of $Q$ versus $\rho$ is called a "fundamental diagram."

The following subsection shows that the evolution equations for particle
CA5s that belong
to certain types can be obtained in the form of max-min-plus expressions
from a fundamental diagram.
\subsection{Type-A}
Let us take particle CA5-34 (Rule 3163536512) as an example. The evolution equation
for particle CA5-34 is given by 
\begin{equation}
 u_j^{n+1} = f(u_{j-2}^n, u_{j-1}^n, u_j^n, u_{j+1}^n, u_{j+2}^n)
\end{equation}
together with the following rule table of $f$.

\begin{center}
{\setlength{\tabcolsep}{1.4pt}\scriptsize
\begin{tabular}{|c||c|c|c|c|c|c|c|c|c|c|c|c|c|c|c|c|}\hline
 $abcde$ & 11111 & 11110 & 11101 & 11100 & 11011 & 11010 & 11001 & 11000 & 10111 & 10110 & 10101 & 10100 & 10011 & 10010 & 10001 & 10000 \\ \hline
 $f(a,b,c,d,e)$ & 1 & 0 & 1 & 1 & 1 & 1 & 0 & 0 & 1 & 0 & 0 & 0 & 1 & 1 & 1 & 1 \\\hline
 $abcde$ & 01111 & 01110 & 01101 & 01100 & 01011 & 01010 & 01001 & 01000 & 00111 & 00110 & 00101 & 00100 & 00011 & 00010 & 00001 & 00000 \\\hline
 $f(a,b,c,d,e)$ & 1 & 0 & 1 & 1 & 1 & 1 & 0 & 0 & 1 & 0 & 0 & 0 & 0 & 0 & 0 & 0 \\\hline
\end{tabular}}
\end{center}
The upper row of the table shows all possible combinations of binary
arguments, and the lower row gives the value of $f$.

From (\ref{HT1}) and (\ref{HT2}), the evolution equation for particle CA5-34 can
be rewritten as
\begin{equation}
 u_j^{n+1} = u_j^n + q(u_{j-2}^n, u_{j-1}^n, u_j^n, u_{j+1}^n) -
  q(u_{j-1}^n, u_j^n, u_{j+1}^n, u_{j+2}^n) \label{PCA5-34}
\end{equation}
together with the following rule table of $q$.
\begin{center}
{\setlength{\tabcolsep}{1.3pt}\small
\begin{tabular}{|c||c|c|c|c|c|c|c|c|c|c|c|c|c|c|c|c|}\hline
$abcd$ & 1111&1110&1101&1100&1011&1010&1001&1000&0111&0110&0101&0100&0011&0010&0001&0000 \\ \hline 
$q(a,b,c,d)$ & 0 & 1 & 1 & 1 &	0 & 0 &	1 & 1 &	0 & 1 &	1 & 1 &	0 & 0 &	0 & 0 \\\hline
\end{tabular}}
\end{center}
\noindent
The upper row of the table shows all possible combinations of binary
arguments, and the lower row gives the value of $q$.

\begin{figure}[htbp]
 \begin{center}
  \begin{overpic}[scale=0.8]{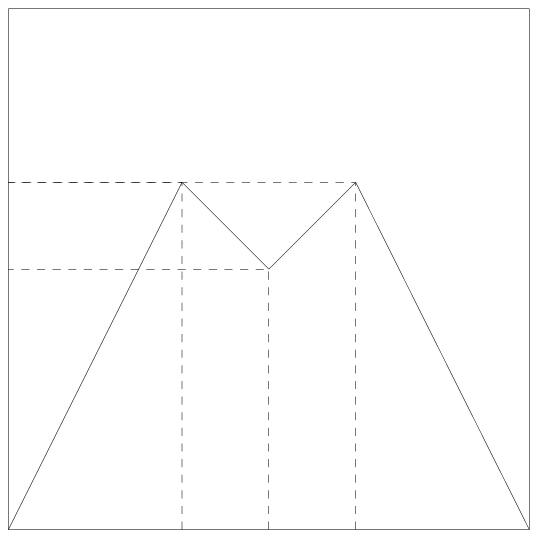}
   \put(5,3){$0$}
   \put(32,3){$1/3$}
   \put(45,3){$1/2$}
   \put(58,3){$2/3$}
   \put(88,3){$1$}
   \put(-2,48){$1/2$}
   \put(-2,61){$2/3$}
   \put(5,88){$1$}
   \put(48,-8){$\rho$}
   \put(-10,48){$Q$}
  \end{overpic}
 \end{center}
 \vspace{1cm}

 \setlength{\tabcolsep}{-5pt}
 \begin{tabular}{cccc}
  \begin{overpic}[scale=0.55]{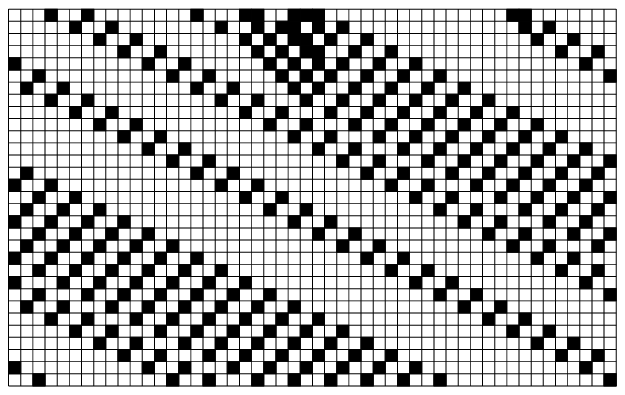}
   \put(57,63){\vector(1,0){10}}
   \put(50,62){$j$}
   \put(5,30){\vector(0,-1){10}}
   \put(3,35){\rotatebox[origin=c]{90}{$n$}}
  \end{overpic}
  &
  \begin{overpic}[scale=0.55]{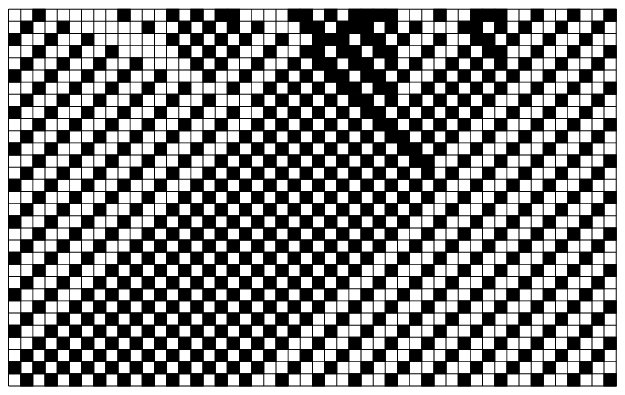}
   \put(57,63){\vector(1,0){10}}
   \put(50,62){$j$}
   \put(5,30){\vector(0,-1){10}}
   \put(3,35){\rotatebox[origin=c]{90}{$n$}}
  \end{overpic}
  &
  \begin{overpic}[scale=0.55]{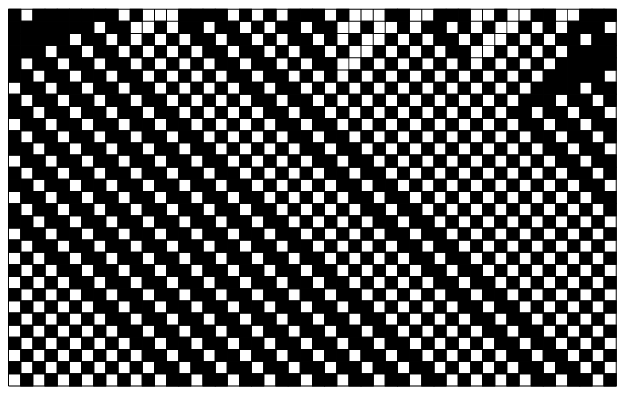}
   \put(57,63){\vector(1,0){10}}
   \put(50,62){$j$}
   \put(5,30){\vector(0,-1){10}}
   \put(3,35){\rotatebox[origin=c]{90}{$n$}}
  \end{overpic}
  &
  \begin{overpic}[scale=0.55]{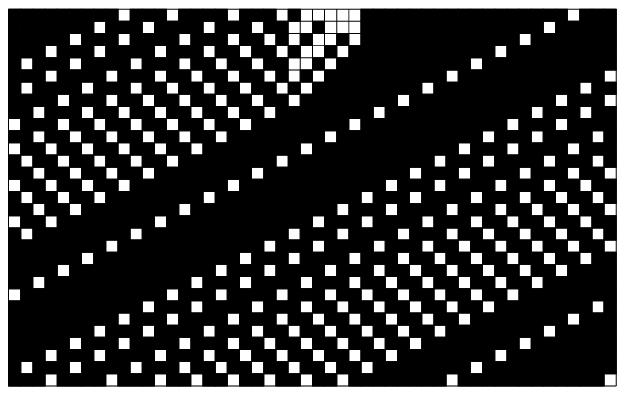}
   \put(57,63){\vector(1,0){10}}
   \put(50,62){$j$}
   \put(5,30){\vector(0,-1){10}}
   \put(3,35){\rotatebox[origin=c]{90}{$n$}}
  \end{overpic}\\
  $\rho = 0.2$ & $\rho = 0.4$ & $\rho = 0.6$ & $\rho = 0.8$\\
 \end{tabular}
 \caption{Fundamental diagram and space-time evolutions of particle CA5-34}
 \label{fig1}
\end{figure}

Employing numerical simulation for (\ref{PCA5-34}), we obtain the
fundamental diagram and examples of evolutions for particle CA5-34 shown in
\Fref{fig1}.

\begin{figure}[htbp]
\begin{center}
\begin{overpic}[scale=0.8]{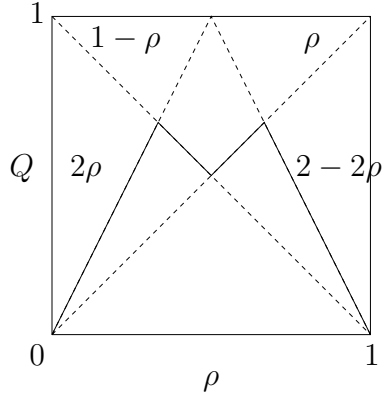}
 \put(5,3){$0$}
 \put(88,3){$1$}
 \put(5,88){$1$}
  \put(15,50){$2\rho$}
 \put(71,50){$2-2\rho$}
 \put(20,82){$1-\rho$}
 \put(73,82){$\rho$}
 \put(48, -2){$\rho$}
 \put(0, 50){$Q$}
\end{overpic}
 \caption{Four linear functions of $\rho$ and the composed piecewise linear function}
 \label{fig2}
\end{center}
\end{figure}

Since the fundamental diagram in \Fref{fig1} is a piecewise linear
curve, the relationship between $Q$ and $\rho$ can be expressed in terms of
the max-min-plus expression,
\begin{equation}
 Q(\rho) = \max(\min(2\rho, 1 - \rho), \min(\rho, 2 - 2\rho)), \label{qrho}
\end{equation}
which is a piecewise linear function composed of four linear functions
$2\rho$, $1-\rho$, $\rho$, and $2-2\rho$. (See \Fref{fig2}.)

\begin{figure}[htbp]
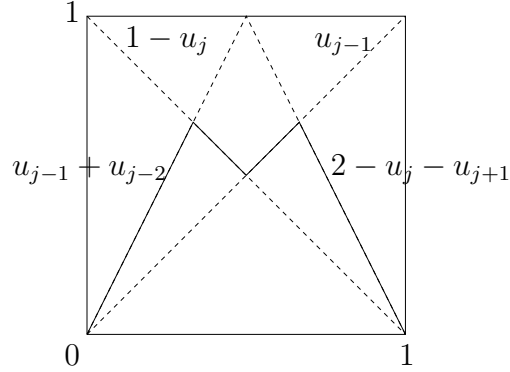

 \begin{center}
\begin{overpic}[scale=0.8]{fig2.eps}
 \put(5,3){$0$}
 \put(88,3){$1$}
 \put(5,88){$1$}
 \put(-8,50){$u_{j-1}+u_{j-2}$}
 \put(71,50){$2-u_j-u_{j+1}$}
 \put(20,82){$1-u_j$}
 \put(67,82){$u_{j-1}$}
\end{overpic}
 \end{center}
 \caption{Obtaining expression for flux from the fundamental diagram}
 \label{fig3}
\end{figure}

Let us consider replacing variables $m\rho$ in (\ref{qrho}) by
\begin{equation}
 m\rho \rightarrow \left\{ \begin{array}{ll}
  \displaystyle \sum_{k=1}^m u_{j-k}^n  & (m > 0) \\[6mm]
  \displaystyle -\sum_{k=1}^{-m} u_{j+k-1}^n  & (m < 0) \\
 \end{array}\right. .
\end{equation}
Then, we obtain
\begin{equation}\fl
 Q(u_{j-2}^n, u_{j-1}^n, u_j^n, u_{j+1}^n) = \max(\min(u_{j-2}^n + u_{j-1}^n, 1 - u_j^n), \min(u_{j-1}^n, 2 - u_j^n - u_{j+1}^n)) \label{flux}
\end{equation}
from (\ref{qrho}). (See \Fref{fig3}.)

Since $Q(u_{j-2}^n, u_{j-1}^n, u_j^n, u_{j+1}^n)$ takes exactly same
values as those of the rule table for $q(u_{j-2}^n, u_{j-1}^n, u_j^n,
u_{j+1}^n)$ of (\ref{PCA5-34}) for all possible combinations of binary
values, (\ref{PCA5-34}) can be rewritten as
\begin{eqnarray}\fl
u_j^{n+1} = u_j^n + \max(\min(u_{j-2}^n + u_{j-1}^n, 1 - u_j^n), \min(u_{j-1}^n, 2 - u_j^n - u_{j+1}^n))\nonumber\\
- \max(\min(u_{j-1}^n + u_j^n, 1 - u_{j+1}^n), \min(u_j^n, 2 - u_{j+1}^n - u_{j+2}^n)).
\end{eqnarray}
This is the max-min-plus expression for particle CA5-34. Introducing the
ultradiscrete Cole-Hopf transformation from $u$ to $F$,
\begin{equation}
 u_j^n = F_{j}^n - F_{j-1}^n,
\end{equation}
we obtain the evolution equation in the form of max-min-plus expressions
for $F$,
\begin{equation}
 F_j^{n+1} =  \min(\max(F_{j-2}^n, F_{j+1}^n - 1), \max(F_{j-1}^n,
  F_{j+2}^n - 2)), \label{PCA5-32f}
\end{equation}
which is composed of linear functions of $F_{j+k}^n\,  (k\in \mathbb{Z})$. Being able to
express evolution equations for $F$ as a combination of linear functions
of $F$ is critical for analyzing the asymptotic behavior of general
solutions of particle CA. We used this fact for the analysis of the asymptotic
behavior of particle CA4 in a previous
study\cite{Takahashi-Matsukidaira-Hara-Feng}.

\begin{figure}[htbp]
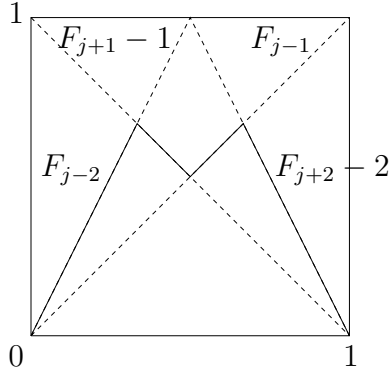

\begin{center}
\begin{overpic}[scale=0.8]{fig2.eps}
 \put(5,3){$0$}
 \put(88,3){$1$}
 \put(5,88){$1$}
 \put(13,50){$F_{j-2}$}
 \put(71,50){$F_{j+2}-2$}
 \put(17,82){$F_{j+1}-1$}
 \put(65,82){$F_{j-1}$}
\end{overpic}
 \caption{Obtaining equations for $F$ from the fundamental diagram}
 \label{fig4}
\end{center}
\end{figure}

Note here that (\ref{PCA5-32f}) can be directly obtained from (\ref{qrho}) by
applying the following replacements. (See \Fref{fig4}.) 
\begin{eqnarray*}
 m\rho + a & \rightarrow \ F_{j-m}^n - a \\
 \max & \rightarrow \ \min \\
 \min & \rightarrow \ \max 
\end{eqnarray*}

It is known that particle CAs allow two different representations: {\it Euler}
representation and {\it Lagrange}
representation. In the Euler
representation, particles are observed at a certain fixed point in space
as dependent(field) variables, while in the Lagrange representation, we
trace each particle and follow its trajectory. Thus, a dependent
variable represents each particle's position in the Lagrange
representation. 

\begin{figure}[htbp]
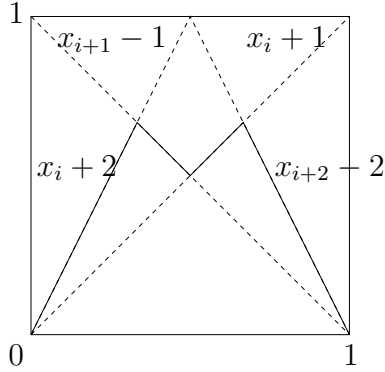

\begin{center}
\begin{overpic}[scale=0.8]{fig2.eps}
 \put(5,3){$0$}
 \put(88,3){$1$}
 \put(5,88){$1$}
 \put(12,50){$x_i+2$}
 \put(71,50){$x_{i+2}-2$}
 \put(17,82){$x_{i+1}-1$}
 \put(64,82){$x_i+1$}
\end{overpic}
 \caption{Obtaining equations for $x$ from the fundamental diagram}
 \label{fig5}
\end{center}
\end{figure}

In previous
studies\cite{Matsukidaira-Nishinari,Matsukidaira-Nishinari2}, we
proposed an Euler-Lagrange transformation for particle CAs by developing the following
transformation formulas for the variable of Euler representation $u_j^n$,
which denotes the number of particles at the $j$th site and $n$th timestep,
and the variable of Lagrange representation $x_i^n$, which denotes the
position of the $i$th particle at the $n$th timestep,
\begin{eqnarray}
& u_j^n = F_j^n - F_{j-1}^n, \\
& F_j^n = \sum_{i=1}^N H(j-x_i^n), \label{transformation}
\end{eqnarray}
where $H(x)$ is the step function defined by $H(x) = 1$ if $x\ge 0$, and
$H(x)=0$ otherwise. Applying the transformation (\ref{transformation})
to (\ref{PCA5-32f}) and using the formulas,
\begin{eqnarray}
& \sum_{i=1}^N H(j-\min(a_i,b_i))=\max(\sum_{k=1}^N H(j-a_i),\sum_{k=1}^N
 H(j-b_i)), \\
& \sum_{i=1}^N H(j-\max(a_i,b_i))=\min(\sum_{k=1}^N H(j-a_i),\sum_{k=1}^N
 H(j-b_i)), \\
& \max(\sum_i H(j-a_i)-m,0) = \sum_i H(j-a_{i+m}),
\end{eqnarray}
where we assume that $a_1 < a_2 < \cdots < a_N$ and $b_1 < b_2 < \cdots
< b_N$, we obtain
\begin{equation}
 x_i^{n+1} =  \max(\min(x_i^n+2, x_{i+1}^n - 1), \min(x_i^n+1, x_{i+2}^n
  - 2)), \label{PCA5-32x}
\end{equation}
which is the Lagrange representation for particle CA5-34 in the form of a max-min-plus
expression. Note here that (\ref{PCA5-32x}) can be directly obtained from (\ref{qrho})
by applying the following replacements. (See \Fref{fig5}.) 
\begin{equation}
 m\rho + a \rightarrow x_{i+a} + m 
\end{equation}

To summarize, the procedure we have performed above is as follows. 

\begin{enumerate}
 \item Employ numerical simulation for particle CA and obtain the fundamental diagram. 
 \item If the fundamental diagram is a piecewise linear curve, construct
       $Q(\rho)$ from the fundamental diagram.
 \item Construct flux $q$ from $Q(\rho)$ and obtain the evolution equation (Euler representation),
       \[
        u_j^{n+1} = u_j^n + q(u_{j-2}^n, u_{j-1}^n, u_j^n, u_{j+1}^n) -
       q(u_{j-1}^n, u_j^n, u_{j+1}^n, u_{j+2}^n).
       \]
 \item Obtain the evolution equation for $F$,
       \[
        F_j^{n+1} = \phi(F_{j-2}^n, F_{j-1}^n, F_j^n, F_{j+1}^n, F_{j+2}^n).
       \]
 \item Obtain the evolution equation for $x$ (Lagrange representation),
       \[
       x_i^{n+1}=p(x_{i-2}^n, x_{i-1}^n, x_i^n, x_{i+1}^n, x_{i+2}^n).
       \]
\end{enumerate}

\begin{figure}[htbp]
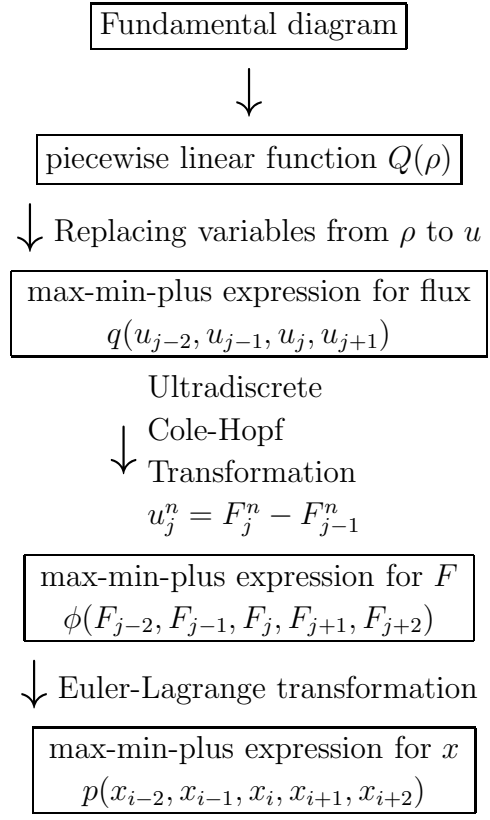

 \begin{center}
 \fbox{Fundamental diagram}
 \medskip

 {\LARGE $\downarrow$}
 \medskip

 \fbox{piecewise linear function $Q(\rho)$}
 \medskip

 {\LARGE $\downarrow$} Replacing variables from $\rho$ to $u$
 \medskip

 \begin{tabular}{|c|}\hline
  max-min-plus expression for flux \\
  $q(u_{j-2}, u_{j-1}, u_j, u_{j+1})$\\\hline
 \end{tabular}
 \medskip

 {\LARGE $\downarrow$} \parbox{3.2cm}{Ultradiscrete Cole-Hopf
 Transformation $u_j^n = F_j^n - F_{j-1}^n$}
 \medskip

 \begin{tabular}{|c|}\hline
  max-min-plus expression for $F$\\
  $\phi(F_{j-2}, F_{j-1}, F_j, F_{j+1}, F_{j+2})$\\\hline
 \end{tabular}
 \medskip

 {\LARGE $\downarrow$} Euler-Lagrange transformation
 \medskip

 \begin{tabular}{|c|}\hline
  max-min-plus expression for $x$\\
  $p(x_{i-2}, x_{i-1}, x_i, x_{i+1}, x_{i+2})$\\\hline
 \end{tabular}
  \caption{Procedure for obtaining evolution equations of type-A
  particle CA5 in max-min-plus expression}
  \label{fig6}
 \end{center}
\end{figure}

Among the 115 rules for particle CA5, there are 17 for which the evolution equations
for $u_j^n$, $F_j^n$, and $x_i^n$ can be obtained in the form of
max-min-plus expressions from fundamental diagrams by using the procedure
described above. These equations are given in Appendix A. In the tables of
the Appendix, $m$ denotes the number of particle CA5-$m$ and $N(f)$ denotes the rule
number defined by Wolfram. Hereafter, we call the 17 rules type-A.

\subsection{Type-B}

Other than type-A, there exist particle CA5 rules for which the fundamental diagram is
a piecewise linear curve, and the evolution equations for $u_j^n$, $F_j^n$, and
$x_i^n$ can be obtained.

\begin{figure}[htbp]
\begin{center}
\begin{overpic}[scale=0.5]{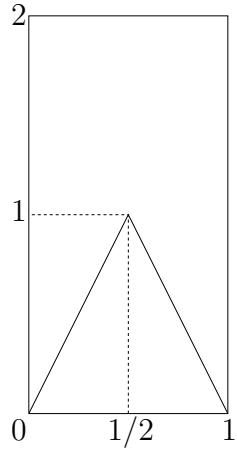}
 \put(2,0){0}
 \put(50,0){1}
 \put(24,0){$1/2$}
 \put(2,50){1}
 \put(2,95){2}
\end{overpic}
 \caption{Fundamental diagram of particle CA5-15}
 \label{fig7}
\end{center}
\end{figure}

Let us consider particle CA5-15 (rule 3099572352). Employing numerical
simulation, we obtain the fundamental diagram for particle CA5-15, as shown in
\fref{fig7}. \fref{fig7} indicates that the piecewise linear curve can be
composed from two linear functions $2\rho$ and $2-2\rho$ as follows.
\begin{equation}
 Q(\rho) = \min(2\rho, 2-2\rho)  \label{PCA5-15qrhox}
\end{equation}
Following the same procedure as in the case of type-A, we obtain a
flux as follows:
\begin{equation}
 q(u_{j-2}, u_{j-1}, u_j, u_{j+1}) = \min(u_{j-2}+u_{j-1},2-u_j-u_{j+1}) \label{PCA5-15qx}
\end{equation}
Evaluating the values of (\ref{PCA5-15qx}) for all possible combinations of
binary values, we obtain the following table.
\begin{center}
{\setlength{\tabcolsep}{1.3pt}\small
\begin{tabular}{|c||c|c|c|c|c|c|c|c|c|c|c|c|c|c|c|c|}\hline
$abcd$ & 1111&1110&1101&1100&1011&1010&1001&1000&0111&0110&0101&0100&0011&0010&0001&0000 \\ \hline 
$q(a,b,c,d)$ & 0 & 1 & 1 & 2 &	0 & 1 &	1 & 1 &	0 & 1 &	1 & 1 &	0 & 0 &	0 & 0 \\\hline
\end{tabular}}
\end{center}
However, the flux of particle CA5-15 is given by
\begin{center}
{\setlength{\tabcolsep}{1.3pt}\small
\begin{tabular}{|c||c|c|c|c|c|c|c|c|c|c|c|c|c|c|c|c|}\hline
$abcd$ & 1111&1110&1101&1100&1011&1010&1001&1000&0111&0110&0101&0100&0011&0010&0001&0000 \\ \hline 
$q(a,b,c,d)$ & 0 & 1 & 1 & 1 &	0 & 1 &	1 & 1 &	0 & 1 &	1 & 1 &	0 & 0 &	0 & 0 \\\hline
\end{tabular}}
\end{center}
and the value of $q(1,1,0,0)$ is found to be different. Thus, 
(\ref{PCA5-15qx}) is not the flux of particle CA5-15. 

\begin{figure}[htbp]
\begin{center}
\begin{tabular}{ccc}
\begin{overpic}[scale=0.5]{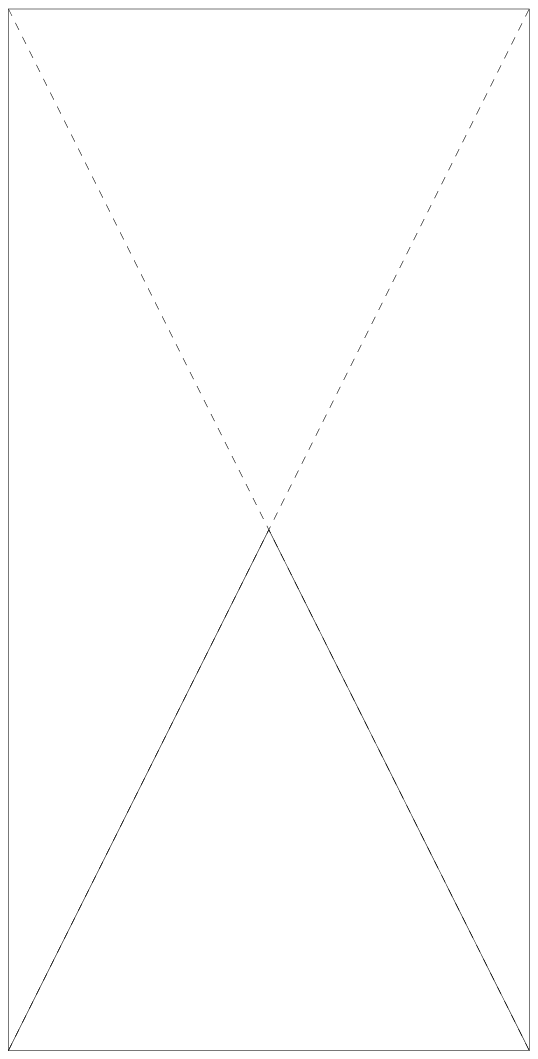}
 \put(2,1){0}
 \put(50,1){1}
 \put(2,50){1}
 \put(2,95){2}
 \put(15,43){$2\rho$}
 \put(34,43){$2-2\rho$}
\end{overpic}
 &
\raisebox{2.6cm}{\Huge $\rightarrow$}
 &
\begin{overpic}[scale=0.5]{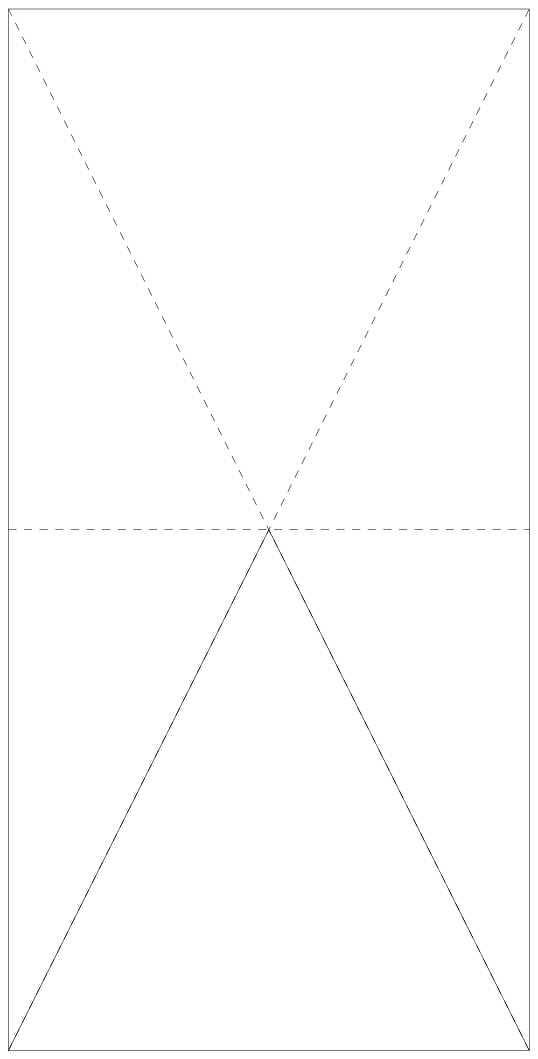}
 \put(2,1){0}
 \put(50,1){1}
 \put(2,50){1}
 \put(2,95){2}
 \put(15,43){$2\rho$}
 \put(34,43){$2-2\rho$}
 \put(16,55){\large 1}
\end{overpic}
\end{tabular}
\end{center}
 \caption{$\min(2\rho,2-2\rho)$ and $\min(1,2\rho,2-2\rho)$}
 \label{fig8}
\end{figure}

To solve this problem, let us suppose that the piecewise linear curve
in \Fref{fig7} is not composed of two but more than
two linear functions. Let us consider
\begin{equation}
 Q(\rho) = \min(1,2\rho,2-2\rho) \label{PCA5-15qrho}  
\end{equation}
instead of (\ref{PCA5-15qrhox}). Note here that (\ref{PCA5-15qrhox}) and
(\ref{PCA5-15qrho}) are identical functions for real values of
$\rho$. (See \Fref{fig8}.)

If we apply variable replacement from $\rho$ to $u$ to the above
expression, we obtain
\begin{equation}
 q(u_{j-2}, u_{j-1}, u_j, u_{j+1}) = \min(1,u_{j-2}+u_{j-1},2-u_j-u_{j+1}) \label{PCA5-15q}
\end{equation}
which gives the following table: 
\begin{center}
{\setlength{\tabcolsep}{1.3pt}\small
\begin{tabular}{|c||c|c|c|c|c|c|c|c|c|c|c|c|c|c|c|c|}\hline
$abcd$ & 1111&1110&1101&1100&1011&1010&1001&1000&0111&0110&0101&0100&0011&0010&0001&0000 \\ \hline 
$q(a,b,c,d)$ & 0 & 1 & 1 & 1 &	0 & 1 &	1 & 1 &	0 & 1 &	1 & 1 &	0 & 0 &	0 & 0 \\\hline
\end{tabular}}
\end{center}
which is identical to the flux of particle CA5-15. Thus, the evolution
equation in the form of a max-min-plus expression for particle CA5-15 is
\begin{equation}\fl
 q_j^{n+1} = q_j^n + \min(1,u_{j-2}^n+u_{j-1}^n,2-u_j^n-u_{j+1}^n) - \min(1,u_{j-1}^n+u_j^n,2-u_{j+1}^n-u_{j+2}^n).
\end{equation}
We can also obtain the evolution equations for $F$ and $x$ as follows.
\begin{equation}\fl
 F_j^{n+1} = \max(F_j^n - 1, F_{j-2}^n, F_{j+2}^n-2)
\end{equation}
\begin{equation}\fl
 x_i^{n+1} = \min(x_{i+1}^n, x_i^n + 2, x_{i+2}^n-2) = x_i^n + \min(2,
  x_{i+1}^n - x_i^n, x_{i+2}^n - x_i^n - 2)
\end{equation}

There are nine rules for particle CA5, of which the evolution equations for $u$, $F$, and $x$
are obtained using the procedure described above. These equations are given in
Appendix B. We call the nine rules type-B.

\subsection{Particle CA5 rules other than type-A and type-B}

We have employed numerical simulations for all 115 particle CA5 rules to obtain
fundamental diagrams, from which we have obtained evolution equations in
the form of max-min-plus expressions for 17 rules of type-A and nine rules
of type-B. There are 89 rules for particle CA5, for whom evolution equations have not yet been obtained. From the results of numerical simulations, the 89
rules are classified into the following two cases.
\begin{itemize}
 \item Although the obtained fundamental diagram is a piecewise linear
       curve, replacing variables from $\rho$ to $u$ does not give the
       correct flux expression for $q$. In addition, the type-B procedure does not seem to work well.
 \item As the obtained fundamental diagram is not a piecewise linear
       curve, we cannot use the procedure of replacing variables from
       $\rho$ to $u$.
\end{itemize}
At this point, it is not clear if there may be other procedures for
obtaining max-min-plus expressions for them.

\section{Particle CA$n$}
We can extend our analysis in the previous section to neighborhood-$n$
CA, i.e., the particle CA$n$ case.

Let us consider the case of $n=6$. One example of particle CA6 is Rule
13755053124876288240. From numerical simulation, we obtain the
fundamental diagram for the rule shown in \Fref{fig6}. 
\begin{figure}[htbp]
 \begin{center}
  \begin{overpic}[scale=0.8]{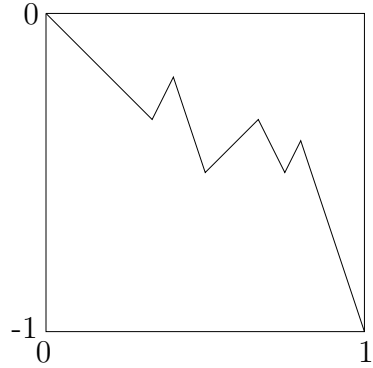}
   \put(5,87){0}
   \put(2,9){-1}
   \put(8,3){0}
   \put(88,3){1}
  \end{overpic}
  \caption{Fundamental diagram for Rule 13755053124876288240}
  \label{fig6}
 \end{center}
\end{figure}
Following the procedure of type-A of particle CA5, we obtain
\begin{eqnarray}\fl
q(u_{j-2}, u_{j-1}, u_j, u_{j+1}, u_{j+2}) = \max(-u_j, \min(u_{j-2}+u_{j-1}-1,1-u_j-u_{j+1}-u_{j+2})),\nonumber\\
\min(u_{j-1}-1,1-u_j-u_{j+1},\min(u_{j-2}+u_{j-1}-2,2-u_j-u_{j+1}-u_{j+2})), \nonumber\\
\end{eqnarray}
\begin{eqnarray}\fl
 F_j^{n+1} = \min(F_{j+1}^n, \max(F_{j-2}^n+1,
  F_{j+3}^n-1),\max(F_{j-1}^n+1,F_{j+2}^n-1),\max(F_{j-2}^n+2,F_{j+3}^n-2)), \nonumber\\
\end{eqnarray}
\begin{eqnarray}\fl
 x_i^{n+1} = \max(x_i^n-1, \min(x_{i-1}^n+2,
  x_{i+1}^n-3),\min(x_{i-1}^n+1,x_{i+1}^n-2),\min(x_{i-2}^n+2,x_{i+2}^n-3)). \nonumber\\
\end{eqnarray}
Although we have not investigated all particle CA6 rules, the analysis
of particle CA5 in
the previous section and the example of particle CA6 above assure us that our
procedure can be applied to particle CA$n$ rules.

\section{Summary}

We have studied particle CA5 and shown that the evolution equations
for type-A and type-B can be obtained in the form of max-min-plus
expressions from a fundamental diagram. The obtained equations have been
transformed into max-min-plus expressions that are composed of linear
functions of $F$ by ultradiscrete Cole-Hopf transformation. Furthermore,
we have obtained the Lagrange representation of the evolution equations.

Although we have not obtained evolution equations for all 115 particle CA5
rules, it is important for us to have been able to introduce a unified
approach using max-plus algebra, together with the fundamental diagram, for
examining particle CAs.

In a previous study\cite{Takahashi-Matsukidaira-Hara-Feng}, we analyzed
asymptotic behaviors of solutions and derived functions $Q(\rho)$ for
particle CA4 mathematically. We have not done this for particle CA5, but
will be able to do so starting from the equations for $F$ obtained in
this study. This problem will be addressed in a future study.

Finally, investigating the generalization to the neighborhood-$n$ case is
important for the future, and we will report on this in a forthcoming paper.

\appendix

\section{Type-A}

\footnotesize
\begin{tabular}{|r|c|l|}\hline
$m$& $N(f)$     & $q(u_{j-2}, u_{j-1}, u_j, u_{j+1})$ \\\hline
 1 & 2863377064 & $\min\left(u_{j-2}+u_{j-1},1-u_j-u_{j+1}\right)$\\\hline
 3 & 2881267852 & $\max\left(-u_j, \min(u_{j-2}+u_{j-1}-1,1-u_j-u_{j+1})\right)$\\\hline
 4 & 2881398914 & $\max\left(-u_j-u_{j+1}, \min(u_{j-2}+u_{j-1}-1,1-u_j-u_{j+1})\right)$\\\hline
 5 & 2881464448 & $\min\left(u_{j-2}+u_{j-1},2-u_{j}-u_{j+1} \right)$\\\hline
33 & 3163470978 & $\max\left( -u_{j}-u_{j+1},\min\left(u_{j-2}+u_{j-1}-1,-u_{j} \right),\min\left(u_{j-1}-1,1-u_{j}-u_{j+1} \right)\right)$\\\hline
34 & 3163536512 & $\max\left(\min\left(u_{j-2}+u_{j-1},1-u_{j}
 \right),\min\left(u_{j-1},2-u_{j}-u_{j+1} \right) \right)$\\\hline
38 & 3167521996 & $\max\left(-u_{j},\min\left(u_{j-2}+u_{j-1}-2 ,1-u_{j}-u_{j+1}\right)\right)$\\\hline
39 & 3167653058 & $\max\left(-u_{j}-u_{j+1} ,\min\left(u_{j-2}+u_{j-1}-1,-u_{j} \right),\min\left(u_{j-2}+u_{j-1}-2,1-u_{j}-u_{j+1} \right)\right)$\\\hline
40 & 3167718592 & $\max\left(\min\left(u_{j-2}+u_{j-1},1-u_{j}\right),\min\left(u_{j-2}+u_{j-1}-1,2-u_{j}-u_{j+1} \right)\right)$\\\hline
50 & 3200487104 & $\max\left(\min\left(u_{j-2}+u_{j-1},1-u_{j}-u_{j+1} \right),\min\left(u_{j-1},1-u_{j} \right),\min\left(u_{j-2}+u_{j-1}-1,2-u_{j}-u_{j+1} \right) \right)$\\\hline
54 & 3203105320 &  $\max\left(\min\left(u_{j-2}+u_{j-1},1-u_{j}-u_{j+1} \right),\min\left(u_{j-2}+u_{j-1}-1,2-u_{j}-u_{j+1} \right) \right)$\\\hline
59 & 3217067968 & $\max\left(\min\left(u_{j-1},1-u_{j}
 \right),\min\left(u_{j-2}+u_{j-1}-1,2-u_{j}-u_{j+1} \right)
 \right)$\\\hline
64 & 3219686184 & $\max\left(\min\left(u_{j-1},1-u_{j}-u_{j+1}\right),\min\left(u_{j-2}+u_{j-1}-1,2-u_{j}-u_{j+1} \right)\right)$\\\hline
65 & 3220209904 & $\max\left(0,\min\left(u_{j-2}+u_{j-1}-1,2-u_{j}-u_{j+1}\right)\right)$\\\hline
68 & 3220996108 & $\max\left(-u_{j}, \min\left(u_{j-2}+u_{j-1}-1,2-u_{j}-u_{j+1} \right)\right)$\\\hline
69 & 3221127170 & $\max\left(-u_{j}-u_{j+1},\min\left(u_{j-2}+u_{j-1}-1,2-u_{j}-u_{j+1} \right)\right)$\\\hline
96 & 3488138024 & $\max(\min(u_{j-1}, 1-u_j^n-u_{j+1}),\min(u_{j-2}+u_{j-1}-1,1-u_j))$\\\hline
\end{tabular}
\bigskip

\noindent
\begin{tabular}{|r|c|l|}\hline
$m$ & $N(f)$ & $\phi(F_{j-2}, F_{j-1}, F_j, F_{j+1}, F_{j+2})$\\\hline
 1 & 2863377064 & $\max\left(F_{j-2},F_{j+2}-1 \right)$\\\hline
 3 & 2881267852 & $\min\left(F_{j+1},\max\left(F_{j-2}+1,F_{j+2}-1 \right)\right)$\\\hline
 4 & 2881398914 & $\min\left(F_{j+2},\max\left(F_{j-2}+1,F_{j+2}-1 \right)\right)$\\\hline
 5 & 2881464448 & $\max\left(F_{j-2},F_{j+2}-2 \right)$\\\hline
33 & 3163470978 & $\min\left( F_{j+2},\max\left(F_{j-2}+1,F_{j+1} \right),\max\left(F_{j-1}+1,F_{j+2}-1 \right)\right)$\\\hline
34 & 3163536512 & $\min\left(\max\left(F_{j-2},F_{j+1}-1 \right),\max\left(F_{j-1},F_{j+2}-2 \right) \right)$\\\hline
38 & 3167521996 &$\min\left(F_{j+1},\max\left(F_{j-2}+2,F_{j+2}-1 \right) \right)$\\\hline
39 & 3167653058 & $\min\left(F_{j+2},\max\left(F_{j-2}+1,F_{j+1} \right),\max\left( F_{j-2}+2,F_{j+2}-1\right) \right)$\\\hline
40 & 3167718592 & $\min\left(\max\left(F_{j-2},F_{j+1}-1 \right),\max\left(F_{j-2}+1,F_{j+2}-2 \right) \right)$\\\hline
50 & 3200487104 & $\min\left(\max\left(F_{j-2},F_{j+2}-1 \right),\max\left(F_{j-1},F_{j+1}-1 \right),\max\left(F_{j-2}+1,F_{j+2}-2 \right) \right)$\\\hline
54 & 3203105320 & $\min\left(\max\left(F_{j-2},F_{j+2}-1\right),\max\left(F_{j-2}+1,F_{j+2}-2 \right)
 \right)$\\\hline
59 & 3217067968 & $\min\left(\max\left(F_{j-1},F_{j+1}-1 \right),\max\left(F_{j-2}+1,F_{j+2}-2 \right) \right)$\\\hline
64 & 3219686184 & $\min\left(\max\left(F_{j-1},F_{j+2}-1 \right),\max\left(F_{j-2}+1,F_{j+2}-2 \right)\right)$\\\hline
65 & 3220209904 & $\min\left(F_{j},\max\left(F_{j-2}+1,F_{j+2}-2\right) \right)$\\\hline
68 & 3220996108 & $\min\left(F_{j+1},\max\left(F_{j-2}+1,F_{j+2}-2 \right) \right)$\\\hline
69 & 3221127170 & $\min\left(F_{j+2},\max\left(F_{j-2}+1,F_{j+2}-2 \right) \right)$\\\hline
96 & 3488138024 & $\min\left(\max\left(F_{j-1},F_{j+2}-1 \right),\max\left(F_{j-2}+1,F_{j+1}-1 \right) \right)$\\\hline
\end{tabular}

\noindent
\begin{tabular}{|r|c|l|}\hline
$m$ & $N(f)$ & $p(x_{i-2}, x_{i-1}, x_i, x_{i+1}, x_{i+2})$ \\\hline
 1 & 2863377064 & $\min(x_i+2, x_{i+1}-2)$ \\\hline
 3 & 2881267852 & $\max(x_i-1,\min(x_{i-1}+2,x_{i+1}-2))$\\\hline
 4 & 2881398914 & $\max(x_i-2,\min(x_{i-1}+2,x_{i+1}-2))$\\\hline
 5 & 2881464448 & $\min(x_i+2,x_{i+2}-2)$\\\hline
33 & 3163470978 & $\max(x_i-2,\min(x_{i-1}+2,x_i-1),\min(x_{i-1}+1,x_{i+1}-2))$\\\hline
34 & 3163536512 & $\max(\min(x_i+2,x_{i+1}-1),\min(x_i+1,x_{i+2}-2))$\\\hline
38 & 3167521996 & $\max(x_i-1,\min(x_{i-2}+2,x_{i+1}-2))$\\\hline
39 & 3167653058 & $\max(x_i-2,\min(x_{i-1}+2,x_i-1),\min(x_{i-2}+2,x_{i+1}-2))$\\\hline
40 & 3167718592 & $\max(\min(x_i+2,x_{i+1}-1),\min(x_{i-1}+2,x_{i+2}-2))$\\\hline
50 & 3200487104 & $\max(\min(x_i+2,x_{i+1}-2),\min(x_i+1,x_{i+1}-1),\min(x_{i-1}+2,x_{i+2}-2))$\\\hline
54 & 3203105320 & $\max(\min(x_i+2,x_{i+1}-2),\min(x_{i-1}+2,x_{i+2}-2))$ \\\hline
59 & 3217067968 & $\max(\min(x_i+1,x_{i+1}-1),\min(x_{i-1}+2,x_{i+2}-2))$ \\\hline
64 & 3219686184 & $\max(\min(x_i+1,x_{i+1}-2),\min(x_{i-1}+2,x_{i+2}-2))$ \\\hline
65 & 3220209904 & $\max(0,\min(x_{i-1}+2,x_{i+2}-2))$ \\\hline
68 & 3220996108 & $\max(x_i-1,\min(x_{i-1}+2,x_{i+2}-2))$ \\\hline
69 & 3221127170 & $\max(x_i-2,\min(x_{i-1}+2,x_{i+2}-2))$ \\\hline
96 & 3488138024 & $\max(\min(x_i+1,x_{i+1}-2),\min(x_{i-1}+2,x_{i+1}-1))$ \\\hline
\end{tabular}

\section{Type-B}

\begin{tabular}{|r|c|l|}\hline
$m$& $N(f)$     & $q(u_{j-2}, u_{j-1}, u_j, u_{j+1})$ \\\hline
 2 & 2881005752 & $\min\left(\max\left(0, u_{j-2}+u_{j-1}-1 \right),1-u_{j}-u_{j+1} \right)$\\\hline
13 & 3099375756 & $\max\left(-u_{j} ,\min\left(0,u_{j-2}+u_{j-1}-1,1-u_{j}-u_{j+1} \right)\right)$\\\hline
14 & 3099506818 & $\max\left(-u_{j}-u_{j+1} ,\min\left(0,u_{j-2}+u_{j-1}-1,1-u_{j}-u_{j+1} \right)\right)$\\\hline
15 & 3099572352 & $\min\left(1,u_{j-2}+u_{j-1},2-u_{j}-u_{j+1}\right)$\\\hline
53 & 3202581216 & $\max\left(\min\left(u_{j-2}+u_{j-1} ,\max\left(0,1-u_{j}-u_{j+1}
 \right)\right),\min\left(u_{j-2}+u_{j-1}-1,2-u_{j}-u_{j+1} \right)
 \right)$\\\hline
63 & 3219162080 & $\max\left(\min\left(u_{j-1} ,\max\left(0,1-u_{j}-u_{j+1}
 \right)\right),\min\left(u_{j-2}+u_{j-1}-1,2-u_{j}-u_{j+1} \right)
 \right)$\\\hline
67 & 3220734008 & $\max\left(\min\left(0,1-u_{j}-u_{j+1} \right),\min\left(u_{j-2}+u_{j-1}-1,2-u_{j}-u_{j+1} \right) \right)$\\\hline
95 & 3487613920 & $\max\left(\min\left(u_{j-1} ,\max\left(0,1-u_{j}-u_{j+1} \right)\right),\min\left(u_{j-2}+u_{j-1}-1,1-u_{j} \right) \right)$\\\hline
98 & 3489185848 &  $\max\left(\min\left(0,1-u_{j}-u_{j+1}\right),\min\left(u_{j-2}+u_{j-1}-1,1-u_{j} \right) \right)$\\\hline
\end{tabular}
\bigskip

\noindent
\begin{tabular}{|r|c|l|}\hline
$m$& $N(f)$ & $\phi(F_{j-2}, F_{j-1}, F_j, F_{j+1}, F_{j+2})$\\\hline
 2 & 2881005752 & $\max\left(\min\left(F_{j},F_{j-2}+1 \right),F_{j+2}-1 \right)$\\\hline
13 & 3099375756 & $\min\left(F_{j+1},\max\left(F_{j},F_{j-2}+1,F_{j+2}-1\right)\right)$\\\hline
14 & 3099506818 & $\min\left(F_{j+2},\max\left(F_{j},F_{j-2}+1,F_{j+2}-1\right)\right)$\\\hline
15 & 3099572352 & $\min\left(F_{j}-1,F_{j-2},F_{j+2}-2\right)$\\\hline
53 & 3202581216 & $\min\left(\max\left(F_{j-2},\min\left(F_{j},F_{j+2}-1\right) \right),\max\left(F_{j-2}+1,F_{j+2}-2\right) \right)$\\\hline
63 & 3219162080 & $\min\left(\max\left(F_{j-1},\min\left(F_{j},F_{j+2}-1\right) \right),\max\left(F_{j-2}+1,F_{j+2}-2\right) \right)$\\\hline
67 & 3220734008 & $\min\left(\max\left(F_{j},F_{j+2}-1\right),\max\left(F_{j-2}+1,F_{j+2}-2 \right)\right)$\\\hline
95 & 3487613920 & $\min\left(\max\left(F_{j-1},\min\left(F_{j},F_{j+2}-1\right) \right),\max\left(F_{j-2}+1,F_{j+1}-1\right) \right)$\\\hline
98 & 3489185848 & $\min\left(\max\left(F_{j},F_{j+2}-1 \right),\max\left(F_{j-2}+1,F_{j+1}-1 \right) \right)$\\\hline
\end{tabular}
\bigskip

\noindent
\begin{tabular}{|r|c|l|}\hline
$m$& $N(f)$ & $p(x_{i-2}, x_{i-1}, x_i, x_{i+1}, x_{i+2})$\\\hline
 2 & 2881005752 & $\min(\max(0,x_{i-1}+2),x_{i+1}-2)$ \\\hline
13 & 3099375756 & $\max(x_i-1,\min(0,x_{i-1}+2),x_{i+1}-2)$ \\\hline
14 & 3099506818 & $\max(x_i-2,\min(0,x_{i-1}+2),x_{i+1}-2)$ \\\hline
15 & 3099572352 & $\min(x_{i+1},x_i+2,x_{i+2}-2)$\\\hline
53 & 3202581216 & $\max(\min(x_i+2,\max(0,x_{i+1}-2)),\min(x_{i-1}+2,x_{i+2}-2))$\\\hline
63 & 3219162080 & $\max(\min(x_i+1,\max(0,x_{i+1}-2)),\min(x_{i-1}+2,x_{i+2}-2))$\\\hline
67 & 3220734008 & $\max(\min(x_i,x_{i+1}-2),\min(x_{i-1}+2,x_{i+2}-2))$\\\hline
95 & 3487613920 & $\max(\min(x_i+1,\max(x_i,x_{i+1}-2)),\min(x_{i-1}+2,x_{i+1}-1))$\\\hline
98 & 3489185848 & $\max(\min(x_i,x_{i+1}-2),\min(x_{i-1}+2,x_{i+1}-1))$\\\hline
\end{tabular}
\bigskip

\ack

This work was supported by JSPS KAKENHI Grant Number 22560068.


\section*{References}
\begin{thebibliography}{99}
 \bibitem{Wolfram}
Wolfram S 1986 {\it Theory and Applications of Cellular Automata} (Singapore: World Scientific)
 \bibitem{Wolfram2}
Wolfram S 2002 {\it A New Kind of Science} (Champaign: Wolfram Media)
 \bibitem{Boccara-Fuks}	Boccara N and Fuk\'s H 1998 \JPA {\bf 31} 6007 
 \bibitem{TTMS} Tokihiro T, Takahashi D, Matsukidaira J and Satsuma J 1996 \PRL {\bf 76} 3247--50
 \bibitem{MSTTT} Matsukidaira J, Satsuma J, Takahashi D, Tokihiro T and Torii M 1997 \PL A {\bf 225} 287--295
 \bibitem{Nishinari-Takahashi} Nishinari K and Takahashi D 1998 \JPA {\bf 31} 5439--50
 \bibitem{Takahashi-Matsukidaira-Hara-Feng} Takahashi D, Matsukidaira J, Hara H and Feng B 2011 \JPA {\bf 44} 135102
 \bibitem{Hattori-Takesue} Hattori T and Takesue S 1991 {\it Physica} D {\bf 49} 295--322
 \bibitem{Matsukidaira-Nishinari} Matsukidaira J and Nishinari K 2003 \PRL {\bf 90} 088701
 \bibitem{Matsukidaira-Nishinari2} Matsukidaira J and Nishinari K 2004 {\it Int. J. Mod. Phys.} C {\bf 15} 507--15
\end{thebibliography}
\end{document}